\documentstyle[12pt, fleqn]{article}
\def\ref{\par\noindent\hangindent=6mm\hangafter=1}
\begin{document}
\baselineskip 8mm
\

\vspace{10mm}

\begin{center}

{\bf A Wavelet Space-Scale-Decomposition Analysis of Structures and

Evolution of QSO's Ly$\alpha$ Absorption Lines}

\bigskip
\bigskip

Jesus Pando and Li-Zhi Fang

\bigskip

Department of Physics, University of Arizona, Tucson, AZ 85721

\end{center}

\newpage
\

\vspace{10mm}

\begin{center}

{\bf Abstract}

\end{center}

A wavelet space-scale decomposition (SSD) analysis of large scale structures
in the universe has been developed. The SSD method of identifying and
measuring
structures in the spatial distribution of objects has been demonstrated. The
position and strength (richness) of the identified clusters can be described
by the corresponding
coefficient of the wavelet transform. Using this technique, we systematically
detected the clustering and its evolution of QSO's Ly$\alpha$ forest lines in
real data and simulated samples. We showed
that the clusters of Ly$\alpha$ absorbers do exist on scales as large as
at least 20 h$^{-1}$ Mpc at significance levels of 2-4 $\sigma$. Independent
data sets show about the same strength distribution of the decomposed clusters.
The number densities of the clusters on scales of 10 - 20 h$^{-1}$ Mpc
are found to evolve in an opposite sense as that
of the lines themselves, i.e. they decrease with redshift. We also showed
that the number density and the strength distribution of
clusters can play an important role in testing or discriminating models, i.e.
it can distinguish real data and simulated samples, which cannot be
discriminated by traditional ways. We used Daubechies 4 and
Mallat wavelets as the bases of the SSD. All above-mentioned
conclusions do not depend on either wavelet basis.

\vspace{3cm}

\noindent{\bf Key words:} Ly$\alpha$ forest - large scale structure
- cosmology

\newpage

\noindent{\bf 1. Introduction}

\bigskip

It is generally believed that the Ly$\alpha$ absorption line forest
in QSO spectra comes from intervening absorbers, or clouds, with neutral
hydrogen column densities ranging from about $10^{13}$ to $10^{17}$
cm$^{-2}$ at high redshifts. The spatial distribution of
the absorption lines in the forests should, in principle, be able to
be used to yield information
concerning the formation and evolution of cosmic clustering on much
larger scales than do galaxies. Since the  Ly$\alpha$ clouds are much
more numerous than QSOs, and since Ly$\alpha$ forest samples
suffer less from selection effects, one can expect that Ly$\alpha$
forest systems are good tracers of matter distributions on
larger scales, and that they might  reveal some aspects of cosmic structure
formation at redshifts from $z \sim$ 1 - 4.

However, the problem of clustering in Ly$\alpha$ forest systems is
seriously controversial. Systematic searches for the physical
clustering of Ly$\alpha$ absorption lines began in the early 1980's.
The first study of the distribution of redshifts in the QSO Ly$\alpha$
forest (Sargent {\it et al.} 1980) concluded that no structures could
be identified. Almost all of the results drawn from two-point correlation
function analysis have failed to detect any significant correlation on
the scale in velocity space from $\sim$ 300 to 30,000 km s$^{-1}$. The
first detection of the correlation of Ly$\alpha$ clouds were made
only on size scales of 50-290 km s$^{-1}$  by Webb (1987), see also
Rauch et al. (1993). In fact, the absence of power in the two-point
correlation function has been claimed as a striking characteristic of
the Ly$\alpha$ forest (Weymann 1993).

On the other hand, a series of works based on other methods have detected
the deviation of Ly$\alpha$ forests from uniform or random distributions
with high confidence. For instance, the distribution
of nearest neighbor Ly$\alpha$ line interval is found to be significantly
different from a Poisson distribution (Duncan, Ostriker, \& Bajtlik 1989;
Liu and Jones 1990;). Crotts (1989) directly
identified a void with comoving size 40 h$^{-1}$ Mpc (where h is the
Hubble constant in units of 100 km/s Mpc) in the Ly$\alpha$ forest of
Q0420-388. Subjecting the same data of Crotts
to a method based on Kolmogorov-Smirnoff (K-S) statistic, Fang (1991)
showed that
Ly$\alpha$ absorbers deviate from a uniform distribution on scales as large
as 30-50 h$^{-1}$ Mpc at  $\sim 3\sigma$ significant level.
Contrary to the results obtained using the two point correlation function,
these studies indicate  that there should exist large scale
structures in Ly$\alpha$ forest.

The difference in the conclusions reached by the different methods is due
mainly to the inefficiency of the two-point correlation function in
detecting large scale
structures. Since to determine the two-point correlation function a good
estimate of mean density of the sample is needed, any statistic based on
the amplitude of the two-point correlation function of objects is
insensitive in detecting structures on size scale $r$ if the uncertainty
in the mean density of the sample is comparable to the mean density
enhancement given by a structure over the size scale $r$. In a word,
two-point correlation function is
not infrared (long wavelength) stable. On the other hand, it is impossible
to accurately determine the mean density from observed samples because
of the lack of information of the object's distributions on scales larger
than the sizes of samples considered. The problem is more severe for the
study of high redshift objects like QSO absorption systems, because the
mean number density of the
lines is redshift-dependent. As Liu and Jones (1990) have
shown, some information on clustering of Ly$\alpha$ clouds, especially on
large scales, is lost in two-point correlation function analysis due to
the effects of the finite size of the sample and the uncertainty of
mean number density of lines. The methods of nearest neighbor line interval
and K-S statistics  do not give information of strength and positions
of individual structures in the spatial distribution of the
Ly$\alpha$ absorbers, as these measures are global.

To overcome this difficulty, a statistic based on the change in
the shape, but not on the amplitude, of the two-point correlation
function has been proposed (Mo et al. 1992a, 1992b, Einasto \& Grasmann 1993,
Deng, Xia \& Fang 1994). This method succeeded in detecting typical
scales in the large-scale structure. It was found that some typical
scales in the distribution of QSO Ly$\alpha$ systems are in good
agreement with that in the distribution of galaxies and clusters of
galaxies (Mo et al. 1992a, b). However, this method, like all others
based on the two-point correlation  function, can only detect the
scales of the clustering, but not the location  of the structures.
Moreover, the choice of bin size or the smoothing length
in the two-point correlation statistics leads to an uncertainty in
detecting correlation scales, especially the binning usually uses a
top-hat function which contains components of all wavelengths in
Fourier space.

In this paper, we propose to investigate this problem by the method of
space-scale-decomposition (SSD) based on discrete wavelet transform.
We choose this method because it has been found to be a perfect mathematic
tool to systematically detect structures on various scales in samples
of turbulence and multi-particle physics (Farge 1992,
Greiner, Lipa \& Carruther 1993). The discrete wavelet SSD possesses a
series of mathematical features including space-scale locality, completeness,
invertibility, orthogonality. These properties guaranteed that the
infrared uncertainty will be avoided. The structures can then be
identified and measured {\it simultaneously} in terms of its scale
and position.

The goals of this paper are two-fold. First, we develop the wavelet SSD
method of identification and description of structures on various scales
in 1-dimension LSS samples. Second, using this technique, we analyze the
structures and its evolution in the spatial distribution of Ly$\alpha$
forest lines. A series of features, which have never been detected
by traditional methods, are found: 1) the clusters of Ly$\alpha$
absorption clouds do exist on scales as large as, at least,
20 h$^{-1}$ Mpc at 2-4$\sigma$ level; 2) the number density of these
clusters show an opposite evolution with that of absorption lines, i.e.
the number of the clusters are decreasing with redshifts; 3) the number
density, strength distribution and evolution of these clusters can
effectively distinguish real data with some linear simulation samples, which
passed all tests before the wavelet SSD analysis.
Therefore, wavelet SSD is not only a good mathematical technique, but
a necessary tool to revealed new physical problems of LSS.

The paper is arranged as follows. In \S 2, the method of wavelet
SSD analysis is presented. \S 3 demonstrated the SSD's identification and
description of structures in simulation samples of Ly$\alpha$
forests. In \S 4 we show the results of systematic detection
of clusters of QSO's Ly$\alpha$ absorption lines in two real data sets.
The evolution of the number density of clusters on various scales, and
its comparison with a dynamical models is studied in \S 5. Finally,
conclusion and discussion is in \S 6.

\bigskip

\noindent{\bf 2. Space-scale decomposition of wavelet transform}

\bigskip

\noindent{\it 2.1 The need for wavelet space-scale decomposition}

Space-scale decomposition (SSD), or multiresolution analysis, is not
new in large scale structure study. Many existing methods for identifying
clusters and groups from galaxy surveys can be classified
as SSD. These include: a) identification of structure simply by {\em eyes},
 b) percolation, c) friend-to-friend algorithm,
d) smoothing by a window function, or filtering technique; e) Fourier
transform on finite domain, etc. Strictly speaking, SSD is a technique
designed for resolving an
arbitrary function simultaneously in terms of its standard variable
(say position) and its conjugate counterpart in Fourier space (in this
case wavenumber) in an efficient manner. Because of this requirement, no
one among the above listed methods is qualified. A complete and
consistent SSD should satisfy all the following conditions.

1. Space-scale locality. By definition, the space-scale decomposed
components should be localized on both physical and scale
(Fourier) space. The basis for the decomposition should be
functions which are concentrated on finite domain and vanish outside a
domain of compact support. This requirement excludes the standard Fourier
transform, because the information content of a distribution
is completely delocalized among all the spectral coefficients.
The methods of smoothing by discontinues functions, such
as the popular window -- the top-hat function, should also be
excluded as there are not localized in Fourier space.

2. Completeness. The basis for the decomposition should be complete
because the goal of SSD is not only to identify special type of structures,
but to decompose samples into objects on all scales, regardless the
strength of the clustering. This requirement excludes, at the very
least, any
method based on identifying structures simply by ``eye.''

3. Invertibility. Like the Fourier transform, the transform from sample space
to
SSD coefficient space should be invertible because we need to re-construct
the distribution
on different scales. This precludes smoothing and passband
filtering techniques, which do not give exact reconstruction formulae
for synthesizing the sample from the smoothed distributions.

4. Orthogonality. This requirement guarantees no mixing between
different space-scale domains. Friends-of-friends, percolation and
related methods are ruled out by this requirement. In fact, even continuous
wavelet transforms should also be ruled out because continuous wavelets
form an overcomplete basis, and their basis functions are not
orthogonal. As a consequence, the continuous wavelet transform of
a random sample shows some correlations that are obviously not
in the sample, but in the wavelet transform coefficients themselves.
Therefore, continuous wavelet transforms used in all previous wavelet
studies of LSS (Slezak, Bijaoui \& Mars 1990; Escalera \& Mazure
1992; Escalera, Slezak \& Mazure 1992; Martinez, Paredes \& Saar 1993)
are not appropriate for SSD.
On the other hand, the discrete wavelet transform allows an orthogonal
projection on a minimal number of independent modes. In this case, all
wavelet coefficients are uncorrelated. Therefore, the
difference between continuous and discrete wavelets is essential,
unlike the case for the Fourier transform, for which the difference of the
continuous and discrete basis is technical.

5. Optimization. The orthogonal basis of wavelet transform are obtained
by (space)
translation and (scale) dilation of one mother function (Meyer 1992;
Daubechies 1992). These translation-dilation procedure allows an optimal
compromise in view of the uncertainty principle. Namely, the wavelet
transform gives very good spatial resolution on small scales, and very
good scale resolution on large scales. The Fourier transform on a
finite domain is not optimized
because it is based on trigonometric functions exhibiting increasingly
many oscillations in a window of constant size. In this case the spatial
resolution on  small scales and the range on large scales are
limited by the size of the window.

Therefore, the discrete wavelet transform SSD should be the best
among existing methods of SSD.

\bigskip

\noindent{\it 2.2 Basic formulae of wavelet SSD}

\bigskip

Any one-dimensional sample of point distributions, like Ly$\alpha$ absorption
lines, in the interval [0-1] with resolution $\Delta x$, can be expressed
as a histogram $f(x)$ with $2^{-J}$ bins
\begin{equation}
f(x)=f^{(J)}(x)=\sum_{k=0}^{2^{J}-1}f_{Jk}\phi_{Jk}^{T}(x)
\end{equation}
where $J$ is an integer to be determined by
\begin{equation}
J = \bmod (|\ln\Delta x|/\ln2) +1,
\end{equation}
and $f_{Jk}$ is the value of $f(x)$ in bin $k2^{-J}\leq x \leq
(k+1)2^{-J}$. $\phi_{Jk}^{T}(x)$ in eq.(1) are given by
\begin{equation}
\phi_{jk}^{T}(x) = \left\{ \begin{array}{ll}
1 & \mbox{for $k2^{-j} \leq x \leq (k + 1) \, 2^{-j}$}\\
0 & \mbox{otherwise.}
\end{array} \right.
\end{equation}
where $0 \leq k \leq 2^j -1$. The superscript $T$ denotes $\phi$ being a
top-hat function.
Obviously, $\phi_{jk}^{T}(x)$ can be re-written as
\begin{equation}
\phi_{jk}^{T}(x)=\phi^{T} \, (2^{j} x - k),
\end{equation}
where $j$, $k$ are integer, $0 \leq k \leq 2^j - 1$, and $\phi^{T}(x)$
the top-hat mother function
\begin{equation}
\phi^{T}(x) = \left\{ \begin{array}{ll}
1 & \mbox{for 0 $\leq$ x $\leq$ 1}\\
0 & \mbox{otherwise.}
\end{array} \right.
\end{equation}
Obviously, $\phi_{00}^{T} = \phi^{T}(x)$. Eq.(4) means that functions
$\phi_{jk}^{T}(x)$ are constructed from mother functions (5) by dilating
a factor 2$^{j}$, and translating a number $k$.

In the function $\phi_{jk}^{T}(x)$, the index $j$ denotes the spatial
scale and $k$ the position. For a given
scale $j$ functions $\phi_{jk}^{T}(x)$ are orthogonal with respect to
the index $k$. Since the resolution in $x$-space is equal to or no finer
than $\Delta x$, $f^{J}(x)$ is the expression of $f(x)$ on finest scale
$J$.

In order to find the distribution $f(x)$ on scale $J-1$, we should
expand $f(x)$ into $\phi_{J-1,k}^{T}(x)$. However, this
expansion cannot simply be found by eq.(1), because $\phi_{J-1,k}^{T}(x)$
are not orthogonal to the finer resolution functions $\phi_{Jk}^{T}(x)$.
To solve this problem, we consider a difference function defined by
\smallskip
\begin{equation}
\psi^{T}(x) = \left\{ \begin{array}{ll}
1 & \mbox{for $0 \leq x \leq 1/2$} \\
-1 & \mbox{for $1/2 \leq x \leq 1$} \\
0 & \mbox{otherwise.}
\end{array} \right.
\end{equation}
Similarly, one can construct $\psi_{jk}^{T}(x)$ by dilating and
translating eq.(6) as
\begin{equation}
\psi_{jk}^{T} = \psi^{T}(2^{j} x - k) = \left\{ \begin{array}{ll}
1 & \mbox{for $k2^{-j} \leq x \leq (k + 1/2) \  2^{-j}$}\\
-1 & \mbox{for $(k+1/2)2^{-j} \leq x \leq (k+1)\ 2^{-j}$}\\
0 & \mbox{otherwise.}
\end{array} \right.
\end{equation}
Functions $\psi_{jk}^{T}(x)$ are orthogonal with respect to {\it both}
indexes $j$ and $k$. For a given $j$, $\psi_{jk}^{T}(x)$ are also
orthogonal to functions $\phi_{jk}^{T}(x)$. $\psi_{jk}^{T}(x)$
and $\phi_{jk}^{T}(x)$ are usually called the wavelet and scaling
functions, respectively.

 From eqs.(4) and (7), we have

\begin{equation}
\begin{array}{ll}
\phi_{j,2k}^{T}(x) &
    = \frac{1}{2} (\phi_{j-1,k}^{T}(x) + \psi_{j-1,k}^{T}(x)),\\
\                 &    \   \\
\phi_{j, 2k+1}^{T}(x) &
   = \frac{1}{2}(\phi_{j-1,k}^{T}(x) - \psi_{j-1, k}^{T}(x)),\\
\                 &    \    \\
\end{array}
\end{equation}
where $0 \leq k \leq 2^{j-1}-1$. Eq.(8) shows that all scaling functions,
$\phi_{jk}^{T}(x)$, can be expressed by wavelets {\it and} scaling functions
on scale $j-1$. For this property, $\psi(x)$ is called father functions.

 From eqs.(1) and (8), it is easy to show that
\begin{equation}
f^J(x)=\sum_{k=0}^{2^{J-1}-1}         f_{J-1,k} \phi_{J-1,k}^{T}(x)
     + \sum_{k=0}^{2^{J-1}-1} \tilde{f}_{J-1,k} \psi_{J-1,k}^{T}(x)
\end{equation}
where the mother function coefficient (MFC), $f_{J-1,k}$, and
father function coefficient (FFC), $ \tilde{f}_{J-1,k} $
are given by
\begin{equation}
\begin{array}{ll}
f_{J-1,k}  & = \frac {1}{2} (f_{J,2k} + f_{J,2k+1}), \\
\          &  \   \\
 \tilde{f}_{J-1,k} & = \frac {1}{2} (f_{J,2k} - f_{J,2k+1}).
\end{array}
\end{equation}
In eq.(9), The term containing the mother functions gives the distribution
$f(x)$ on scale $J-1$,
and the term containing the father functions contains the
information of the difference
between scales of $J$ and $J-1$. Since $\psi_{J-1,k}^{T}(x)$ are
orthogonal to $\phi_{J-1,k}^{T}(x)$, the mother function term is not mixed
with any components on scales $J$. Therefore, one can safely describe the
distribution $f(x)$ on scale $J-1$ by
\begin{equation}
f^{J-1}(x)=\sum_{k=0}^{2^{J-1}-1}f_{J-1,k} \phi_{J-1,k}^{T}(x)
\end{equation}
One can repeat this procedure to find the distribution on scale $J-2$
from scale $J-1$. Thus, the distribution $f(x)$ can be decomposed
into $f^j(x)$ with  $0 \leq j \leq J$. For scale $j$, the distribution
$f^j(x)$ is totally determined by the MFC $f_{j,k}$.

For largest scale $j=0$, we have $f^0(x)= f_{0}^{0} \phi_{00}^{T}(x)$,
and
\begin{equation}
f^{J}(x) = f_{0}^{0} \phi_{00}^{T}(x) + \sum_{j=0}^{J-1} \,
\sum_{k=0}^{2^{j}-1}  \tilde{f}_{jk} \psi_{jk}^{T}(x)
\end{equation}
Obviously, the MFC  $f_{0}^{0}$ now is simply the mean density of points
on interval $[0,1]$. Since functions $\psi_{jk}^{T}(x)$ are orthogonal
with respect to $j$ and $k$, the FFCs can be calculated by
\begin{equation}
\tilde{f}_{jk} = 2^{j} \int f(x)\psi_{jk}^{T} dx
\end{equation}
MFC at various scales can be found from $f_{0}^{0}$ and
FFCs $\tilde{f}_{jk}$.

All the above discussions are based on the top-hat wavelet. As we
mentioned in the last section, the to-hat function (5) and difference
function (6) are discontinues, and so are not good for SSD. Nevertheless,
many formulae developed above still hold for all wavelet transforms. For
instance, eq.(13) should be fundamental for any mother functions.
In the mid-80's to early 90's there was a great deal of
work in trying to find a continuous
basis that was well localized in Fourier space (Daubechies et al.
1986, Daubechies
1990, Mallat 1989, Mallat \& Zhong 1990, Meyer 1986). Specifically,
Daubechies (1988)
constructed several families of wavelets and scaling functions which
are orthogonal, have compact support and are continuous. The wavelet
function $\psi_{jk}(x)$ and the scaling function $\phi_{jk}(x)$ are
defined as
\begin{eqnarray}
\phi(x) & = & \sum_{m} c_{m} \, \phi(2x -m) \\
\psi(x) & = & \sum_{m} (-1)^{m} c_{1-m} \; \phi(2x -m)
\end{eqnarray}
where the coefficients $c_{m}$ must satisfy proper conditions (Daubechies
1988). If the nonzero coefficients $c_m$ are taken to be $c_0=c_1=1$,
we have the top-hat scaling and wavelet (5) and (6). The simplest wavelet
function $\psi(x)$ which is dually localized in both $x$-space and
Fourier-space is given by filter $c_0=(1+\sqrt 3)/4$, $c_1=(3+\sqrt 3)$,
$c_2=(3-\sqrt 3)/4$ and $c_3=(1-\sqrt 3/4)$. It is often called wavelet D4.
In our SSD analysis, the decomposition was mainly done by the D4,
but we also check the reliability of the SSD results by using wavelets
of D12, D20 and Mallet (1989).

\bigskip

\noindent{\bf 3. A Demonstration of structure identification by wavelet SSD}

\bigskip

In order to demonstrate the method of wavelet SSD, we did a SSD analysis of
simulation samples of Ly$\alpha$ forests covering redshift range from
1.7 to 4.1 (Bi 1993; Bi, Ge \& Fang 1994, hereafter BGF). The density
field in this simulation are generated as
Gaussian perturbations with linear power spectrum given by cosmological
models of the standard cold dark matter (SCDM), the cold plus hot dark
matter (CHDM), and the low-density flat cold dark matter (LCDM).
Within a reasonable range of $J_\nu$, the UV background radiation
at high redshift, the sample of LCDM model is found to be in good agreement
with observational features including 1) the number density of
Ly$\alpha$ lines and its dependencies on redshift and equivalent width;
2) the distribution of equivalent widths and its redshift dependence; 3)
two-point correlation function; and 4) the Gunn-Peterson effect.

Since the perturbation spectrum used for the simulation is not white noise,
the distribution of Ly$\alpha$ absorption lines in BGF samples should contain
large scale structures. However, the two point correlation function of
lines in redshift (or velocity) space detected nothing from these samples.
A typical result of the two-point correlation function of a BGF sample
best fitting
observations is plotted in Figure 1. As is the case with observations, the
simulated samples showed no power of line-line correlations on scales of
about 100 km s$^{-1}$ to 2000 km s$^{-1}$. These results clearly
show that the two-point correlation function sometimes is ineffective in
detecting structures on large scales.

We then subjected the best fitting sample of the BGF data to a SSD analysis
by the D4 wavelet. First, we formed one dimensional distribution $f(x)$
of Ly$\alpha$ lines by writing each sample in BGF into histogram with
bins of $\bigtriangleup z = 0.0025$, which was about the resolution
with which the data was produced.  We then generated 100 random samples
for each simulation sample. To consider the influence of evolution of
the number of lines, the total number of lines and the number of lines
within a given red-shift interval (say, $\bigtriangleup z = 0.4$) of the
random samples are chosen to match the parent
distribution. We calculated the FFC $\tilde{f}_{jk}$ and MFC $f_{jk}$
for both the BGF sample and the random data. In fact, these amplitudes
can easily be obtained via linear transformations from data space into
wavelet space by a wavelet transformation matrix (Press et al. 1991).

Using MFCs, one can reconstruct the density field $f^j(x)$ on the scale
being considered. Figure 2 shows a result of reconstructing a BGF sample
on scales $j= J-1, J-2$ and $J-3$, where $J$ the finest (resolution) scale,
and $J-1, \ J-2, \ J-3$ correspond to scales (in comoving space) of about
5, 10 and 20 h$^{-1}$ Mpc, respectively. In the reconstructed field the
density is sometimes negative (see Figure 2) because
the mother function of D4 are somewhere is negative. For structure
identification we don't really need a reconstructed field, but only the
MFCs. This shows that the wavelet reconstruction is not
simply a smoothing technique.

Since the MFCs, $f_{jk}$, describes the strength of
density field at position $k$ and on scale $j$, we can identify the clusters
by calculating the MFC's difference between simulated sample and random
data. Figure 3
shows a part of the result of the difference of MFCs between a BGF sample
and random distribution. The error bars are 1$\sigma$ around the average of
the MFC's differences given by 100 random sample. From Figure 3, one
can easily pick out the peaks, each of which is described
by position ($k$) and height of the coefficient difference. The clusters
of Ly$\alpha$ absorption lines are then identified as these peaks. The scale
of these clusters is given by $j$, the
location of clusters is shown by the position of peaks,
and the strength or richness of the clusters can be measured by the
height/$\sigma$.

Figure 4 shows the total number $N(>R)$ of clusters with strength larger
than 2$\sigma$ on scales $J-1, \  J-2, \ \mbox{and}, \  J-3$. To test this
identification we analyzed 20 BGF
simulation samples with the same cosmological parameters. The error
bar in Figures 4 is given by the average among the 20 simulated
samples. One can conclude that the structures in the line
distribution have been systematically detected on scales $j-1, \ j-2, \
\mbox{and} \ j-3$. The identified structures are at significance
levels of 2-4
$\sigma$. It is clear that the wavelet SSD has accomplished what
the two point correlation function could not. The function $N(>R)$
describes the strength (or richness) distribution of the identified
clusters.

\bigskip

\noindent{\bf 4. Clusters of Ly$\alpha$ absorbers}

\bigskip

Having had the wavelet SSD successfully analyze the simulated BGF data, we
now turn to identifying the clusters of Ly$\alpha$ absorbers from real
data. We
should first point out that the word "clusters" used here does not imply
that they are virialized and gravity-confined systems. SSD identified
clusters are not defined by dynamical features like clusters of galaxies,
but only by density distributions. This is not a weakness, but probably
an advantage of wavelet SSD.

Recent measurements have found that the size of the Ly$\alpha$ clouds
at high redshift is unexpectedly as large as 100 - 200 h$^{-1}$ Kpc,
and their velocity dispersion is unexpectedly as low as $\sim$ 100 km
s$^{-1}$ (Bechtold et al. 1994, Dinshaw et al. 1995). These results
cannot be matched with the pictures of pressure equilibrium and virialization.
For instance, if the Ly$\alpha$ clouds with such large size are well
gravitationally confined, the Press-Schechter theory shows that their column
density should be equal to or larger than $10^{17}$ cm$^{-2}$ (Mo,
Miralda-Escud\'e \& Rees 1993). Clouds with large size and low column
densities are not completely gravity-confined. Therefore, the Ly$\alpha$
clouds are probably neither virialized nor completely gravity-confined,
but given by pre-collapsed areas in the density field.

Therefore, identification of dynamically pre-collapsed clusters (or dense
areas) is important, especially for understanding structure formation at
high redshifts. The wavelet SSD would be able to uniformly identify
clusters with various strength, i.e. at various evolutionary stage.

\bigskip

\noindent{\it 4.1 Samples and identification}

\bigskip

We look at two data sets of the Ly$\alpha$ forests. The first was compiled
by  Lu, Wolfe and Turnshek (1991, hereafter LWT). It contains totally
$\sim$ 950 line from the spectra of 38 QSO that exhibit neither broad
absorption lines nor metal line systems. The second is from Bechtold
(1994), which contains a total $\sim$ 2800 lines from 78 QSO's spectra,
in which 34 high redshift QSOs were observed at moderate resolution.
In our statistics, the effect of proximity to $z_{em}$ has been considered.
All lines with
$z \geq z_{em} - 0.15$ were deleted from our samples. As with other data
sets, no power of two point correlation function was detected from
these two compiled samples on large scales.

We treated the data as discussed in the previous section. We assumed
$q_{0} = 1/2$, so the distance of an absorber at redshift $z$ is given
by $d = 2(c/H_{o}) [1 - (1+z)^{-1/2}]$. The samples range from a comoving
distance of about 2,500 $h^{-1}$Mpc to 3,300 $h^{-1}$Mpc. Each QSO's
spectrum was analyzed individually, i.e. for each QSO 100 random trials
matching the line numbers in each redshift range of the parent sample
were generated. Similar to Figure 3, Figure 5 shows a part of the MFC
difference for the forest of QSO-0237. The errors are also the 1$\sigma$
found from the average over the 100 subtractions between MFCs of real
data and random samples.
Figure 5 is typical for all QSO's analyzed. It is interesting to note
from Figure 5 that the $j=J-1$
clusters shown around 2465-2480 h$^{-1}$ Mpc also appear as $j=J-2$
and $j=J-3$ clusters at the same place. That is this structure appears at
all three resolution scales, $j=J-3$,
$j=J-2$ and $j=J-1$. Differently, the structure appearing at 2505-2520 Mpc
only appears on the scales $j= J-1$ and $j= J-2$, but not on larger
scales.

\bigskip

\noindent{\it 4.2 Number and strength of clusters}

\bigskip

In Figures 6, 7 and 8 we show the total number $N(>R)$ of clusters
detected with the strength $ R> 2\sigma $ for both the LWT and Bechtold
data.
It is interesting to note from Figures 6 and 7 that the two independent
data sets show statistically the same strength distributions of the clusters
consisting of $W>0.32\AA$ lines.
However, comparing these results with the BGF's clusters (Figure 4),
the drop
of $N( > R)$ of real data is obviously slower than for the simulated
BGF data. That is,
the abundance of rich clusters in real data is higher than simulated
samples. From Figures 4, 6 and 7, one can find that among all $j=J-1$
clusters of $> 2\sigma$, the abundances of $R >3.5 \sigma$ clusters
is 7\% for BGF sample, 23\% for LWT data, and 28\% for Bechtold data.
The difference of $R >4 \sigma$ cluster abundance between real and
simulated sample is more remarkable.
Almost no $ R > 4.5 \sigma$ clusters are detected in BGF samples,
while they do exist in the real data. This indicates that in linear
simulations the rich clusters are underestimated.

As very well known, the spatial clustering of Ly$\alpha$ absorbers
can be smeared out by peculiar motion of the absorbers. The influence
of peculiar velocity is difficult to estimated because the velocity
distributions of absorbers is not clearly understood.  However, on
the scales equal to or larger than about 5 h$^{-1}$ Mpc, the influence
of peculiar motions should be negligible.  Therefore, the identified
clusters on scales $j=J-1$, $j=J-2$ and $j=J-3$ should be more reliable
than the original lines as a tracers of matter distribution on large
scales.

In order to test the stability of the number of the identified clusters
with respect to the choice of wavelets, we repeated the SSD analysis
by using the Mallet wavelet basis, which is more soft than the D4. Figure
9 shows the number of the identified
($>2 \sigma$) clusters from the LWT data. Comparing Figure 6 and 9,
one finds that the total number of the $R > 2\sigma$ clusters on $j=J-1$
and $j=J-2$ is almost identical for both wavelets. The total
number of structures at
$j=J-3$ shows a slight difference, but is of no statistical significance.

\bigskip

\noindent{\bf 5. Redshift-dependence of clusters}

\bigskip

Using the samples of clusters identified from the wavelet SSD, one can
study all aspects of interest for large scale structure formation: the
number density, correlation function, their redshift-dependence
(evolution), scale- and strength-dependence etc. That is, SSD opens
new fields to compare observations with models. We can conduct the
testing and discriminating of models by a scale-to-scale confrontation.
In this paper, as an example, we only show a scale-to-scale study of
the number density and its redshift-dependence of the clusters of
Ly$\alpha$ absorbers.

\bigskip

\noindent{\it 5.1 Evolution of number density}

\bigskip

It is generally believed that the number of Ly$\alpha$ forest lines
increases with redshift. The redshift-dependence of the number density of
lines with rest equivalent width $W$ greater than a threshold
$W_{th}$ can be described as
\begin{equation}
\frac{dN}{dz}=\left(\frac{dN}{dz}\right)_0(1+z)^{\gamma} \ ,
\end{equation}
where $(dN/dz)_{0}$ is the number density extrapolated to zero redshift,
and $\gamma$ the index of evolution. If the absorbers distribution is
comoving in a flat universe, the number density should be
$(dN/dz) \propto (1+z)^2/[\Omega(1+z)^3 + \lambda]^{1/2}$, where
$\lambda$ is the cosmological constant. The deviation of $dN/dz$ from
the comoving curve implies an evolution of the population of
Ly$\alpha$ clouds. In the case of $\lambda=0$, $\gamma >0.5$ implies
that the number of Ly$\alpha$ clouds increases with redshift, and
$\gamma<0.5$ decreases with redshift. The index $\gamma$ is found
to depend on $W_{th}$: in general
the larger $W_{th}$, the higher $\gamma$.

The simple power law, eq.(16), cannot cover the entire redshift range
being examined. The parameters $(dN/dz)_0$ and $\gamma$ are found
to be different at different redshift ranges. Although the $dN/dz$ given
by different groups showed a common evolutionary trend, there
are differences in amplitude $(dN/dz)_0$ by a factor of about 30\%.
LWT (1991) showed that $(dN/dz)_0 \simeq 3$ and $\gamma = 2.75 \pm 0.29$
for lines with $W\ge W_{th}=0.36\AA$ and in the redshift range
$1.6 < z < 4$. Bechtold (1994) found $\gamma = 1.89 \pm 0.28$ for
$W_{th}=0.32\AA$ and $\gamma=1.32\pm0.24$ for $W_{th}=0.16\AA$.
This smaller value seems to be consistent with the low-redshift results
from the Hubble Space Telescope (Morris et al. 1991; Bahcall et al.
1991.) It would better to directly compare the curves of
$dN/dz \,- \,z$, not the parameters $(dN/dz)_0$ and $\gamma$.

We analyzed the evolution of the number density of clusters on scales
$j=J-1$, $j=J-2$ and $j=J-3$. The results are plotted in Figure 10,
in which 10a is
for LWT and  10b for Bechtold data, both the width is taken to be
$>0.32\AA$. The $dN/dz$ vs $z$ curves of Ly$\alpha$ lines are
the same as the LWT's and Bechtold's original results. Namely, both
show that the
number density of Ly$\alpha$ lines increases with redshift.
However, the number densities of $j=J-1, J-2, J-3$ clusters show
an opposite evolution, i.e. they are decreasing with redshifts.
This result indicates that, along with the possible
decline of Ly$\alpha$ lines, the large scale structures traced by
Ly$\alpha$ lines were growing during the era $4 > z >2$. Obviously,
this evolutionary scenario would not be able to be revealed without
a powerful SSD measurement.

Because we lack sufficient data to provide an accurate analysis of
$dN/dz$ of clusters at high redshift, we are reluctant as yet
to quantify the index $\gamma$ for the clusters on different scales.
Nevertheless, these opposite evolutions should be statistically
significant because the two independent data sets (LWT and Bechtold)
showed the same amplitude of $dN/dz$, and the same trend of the
evolution over the entire redshift range $2< z<4$.

\bigskip

\noindent{\it 5.2 Testing model by scale-decomposition}

\bigskip

Let us go back to the simulation samples of BGF. We did the same
analysis as in the previous section for the best BGF samples.
The result is shown in Figure 11, in which 11A is from BGF sample and
11B from Bechtold data with the same width threshold as BGF,  $W>0.16\AA$.
First, Figure 11 shows that the
$dN/dz$ vs $z$ curve of Ly$\alpha$ lines themself of BGF sample
is in good agreement with observation. This is one reason we say that
this BGF sample is the best. The evolution of the number densities of
$j=J-1,\  J-2, \ J-3$ clusters also show the same trend as real data:
$dN/dz$ is decreasing  with redshift.

However, the number densities given by simulated sample are remarkably
different from the real data. The values of $dN/dz$ of $J-1, \ J-2,
\ J-3$ clusters for BGF are less than the observational results by
a factor of 5-10. Considering
both LWT and Bechtold have about the same number density, one cannot
explain the difference between BGF and Bechtold as the uncertainty
of the current observation. This result once again suggests that the
linear simulation underestimated the clustering on large scales.
In a word, the number densities of $j=J-1, J-2, J-3$ clusters are
effective tools to test models which have passed all tests before
the SSD analysis.

\bigskip

\noindent{\bf 6. Conclusion}

\bigskip

The clustering and evolution of Ly$\alpha$ absorption lines have
been systematically detected by wavelet SSD. It has been shown that
the clusters of Ly$\alpha$ absorbers do exist on scales as large as
at least 20 h$^{-1}$ Mpc at significance level of 2-4 $\sigma$. We
found that the evolution of the number
densities of the clusters is opposite from the lines themself, i.e.
decreasing with redshift. This is probably the first time any one has seen
this phenomenon. We also showed that the number density of the
clusters, and the strength distribution of the clusters can effectively
distinguish
real data with some models, which can not be distinguished from
observational features without a wavelet SSD description.

Therefore, the wavelet SSD is an efficient, fast and reliable way
of detecting structure where other traditional methods have failed.
It provides a vehicle for discovering physics at the dimension of
scales. It open the window to study the scale-dependence of various
features of clustering. For instance, comparing Figure 10B and 11B, one
find that the number densities of $W>0.16\AA$ clusters are higher than
that of $W>0.32 \AA$ clusters, this scale-dependence implies
that the formation of structure in the universe had undergone a
biased clustering with respect to the rest frame equivalent widths of
absorbers.

It is not accidental that the wavelet SSD analysis shows its unique role of
measuring LSS, because the clustering in the universe is probably
multi-scaled. According to the standard picture of the structure
formation, the initial perturbations are scale-free. The successively
non-linear evolution destroyed the perfect scaling, and the density field
was no longer  described as a Gaussian superposition of plane waves.
Therefore, in terms of localized physical and Fourier spaces, the
clustering in the universe would be multiscaled, i.e. the LSS is a
superposition of coherent structures with various scales. It has been
known that in order to fit with galaxies distribution the spectrum of
perturbations in the present universe should contain a set of parameters
(e.g.
Bardeen et al. 1986), each of which should, in principle, correspond
to a scale in the clustering.

Observations also showed the multi-scales. In Ly$\alpha$ forests, at
least three scales have been mentioned: 1. 40 h$^{-1}$ Mpc of a void
Crotts (1989); 2. 30-50 h$^{-1}$ Mpc from K-S statistic (Fang 1991);
3. 80, and even 120 h$^{-1}$ Mpc from  typical scale analysis (Mo, et
al. 1992a, b). In the same redshift range, QSO's clustering have also been
shown to be multi-scaled. Two correlation scales of
$r_0 \sim  6 \ $h$^{-1}$Mpc (Boyle \& Mo 1992), and 13 h$^{-1}$Mpc
(Bahcall \& Chokshi 1991) were found from the two-point correlation
function, and clustering scales of $r\leq 30\ h^{-1}$Mpc and
$\sim$ 100 $h^{-1}$Mpc have also been detected by the statistics of
average two-point correlation function (Mo \& Fang, 1993) and typical
scale analysis (Mo, et al. 1992a,b; Einasto \& Grasmann 1993,
Deng, Xia \& Fang 1994), respectively. Some differences among
these scales may come from the methods used in the analysis, but
it may be difficult to explain all these results by one scale in
clustering.

One can concludes that scale-decomposition is not only mathematically
convenient,
but also physically necessary. We believe that the wavelet SSD of LSS is
a vast new area for exploration. For instance, one can, at least, address
the following topics, which would not be able to be reached without a
qualified SSD:
1.) The scale-dependence of correlations; 2.) non-Gaussianity and the
strength distribution of scale-decomposed clusters; 3.) scale-scale
interaction, i.e. the interaction between structures on different
scales. We will present the results concerning these questions in
succeeding papers.

\bigskip

Both authors wish to thank Drs. Bi, Carruthers, Lipa, and Mo for
insightful conversations.

\newpage

\noindent{\bf References}

\bigskip

\ref Bahcall, J.N., Jannuzi, B.T., Schneider, D.P., Hartig, G.F., Bohlin, R.,
\& Junkkarinen, V. 1991, ApJ,  377, L5

\ref Bahcall, N.A. \& Chokshi, A. 1991, ApJ, 380, L9

\ref Bardeen, J., Bond, J.R., Kaiser, N. \& Szalay, A.S. 1986, ApJ, 304, 15

\ref Bechtold, J. 1994, ApJSS, 91, 1.

\ref Bechtold, J., Crotts, P.S., Duncan, R.C. \& Fang Y. 1994, ApJ,
437, L83

\ref Bi, H.G. 1993, ApJ, 405, 479

\ref Bi, H.G., Ge, J., Fang, L.Z. 1994, BAAS, 26, 1331

\ref Boyle, B.J. \& Mo, H.J. 1992, MNRAS, 260, 952

\ref Chu,Y.Q., Fang, L.Z. 1987 in {\it Observational Cosmology},
eds. A. Hewitt, G. Burbidge, L.Z. Fang, Reidel, Dordrecht, P. 627

\ref Crotts, A.P.S. 1989, ApJ, 336, 550

\ref Daubechies,I., Grossmann,A., Meyer, Y., 1986 J. Math. Phys.
 27, 1271

\ref Daubechies, I. 1988, Comm. on Pure and Applied Mathematics,
41, 909

\ref Daubechies, I. 1990, IEEE Trans. Inf. Theory, 36, 961

\ref Deng, Z.G., Xia, X.Y. \& Fang, L.Z. 1994, ApJ, 431, 506

\ref Dinshaw, N., Foltz, C.B., Impey, C.D., Weymann, R. \& Morris,
S.L. 1995,  Nature, in press

\ref Duncan, R.C., Ostriker, J.P. \& Bajtlik, S. 1989, ApJ, 345, 39

\ref Einasto, J. \& Grasmann, M. 1993, ApJ, 407, 443

\ref Escalera, E., \& Mazure, A. 1992, ApJ, 388, 23

\ref Escalera, E., Slezak, E. \& Mazure, A. 1992, A\&A, 264, 379

\ref Fang, L.Z. 1991, A\&A, 244, 1

\ref Farge, M. 1992, Ann. Rev. Fluid Mach., 24, 395

\ref Greiner, M., Lipa, P., \& Carruthers, P. 1993, HEPHY-PUB 587/93

\ref Lu, L., Wolfe, A.M., \& Turnshek, D.A. 1991, ApJ, 367, 19

\ref Liu X.D., Jones B.J.T. 1990, MNRAS, 242, 678

\ref Mallat, S. 1989, Trans. Am. Math. Soc., 315, 69

\ref Mallat, S., Zhong, S. 1991, {\it Wavelets and Their
Applications\/}, Boston, Jones \& Bartlett.

\ref Martinez, V.J., Paredes, s. \& Saar, E. 1993, MNRAS, 260, 365

\ref Meyer, Y. 1987, {\it Wavelets, Time-Frequency Methods and
Phase Space},  21

\ref Mo, H.J. \& Fang, L.Z. 1993, ApJ, 410, 493

\ref Mo, H.J., Deng, Z.G., Xia, X.Y., Schiller, P., \& B\"orner,
    G. 1992a, A\&A, 257, 1

\ref Mo, H.J., Xia, X.Y., Deng, Z.G., B\"orner, G. \& Fang, L.Z.
    1992b, A\&A, 256, L23

\ref Morris,S.L.,Weymann, R.J., Savage, B.D., \& Gilliland, R.L., 1991,
ApJ,  377, L21.

\ref Press, W.H., Flannery, B.P., Teukolsky, S.A. \& Vetterling, W.T. 1992,
{\it  Numerical Recipes\/}, New York;  Cambridge University Press

\ref Sargent, W.L., Young, P.J., Boksenberg, A., \& Tytler, D. 1980,
ApJS, 42, 41

\ref Slezak, E., Bijaoui, A. \& Mars, G. 1990, A\&A, 227, 301

\ref Rauch, M., Carswell, R.F., Chaffee, F.H., Foltz, C.B., Webb, J.K.,
Weymann, R.J., Bechtold, J. \& Green, R. F. 1992, ApJ, 390, 387

\ref Webb, J.K. 1987, in {\it  Observational Cosmology}
e. A. Hewitt., G. Burbidge, \& L.Z. Fang.

\ref Weymann, R.J. 1992, in {\it The Environment and Evolution of
Galaxies\/}, ed. Shull, J.M. \& Thronson, H.A. Jr. 214 (KLumer
Academic Publisher).

\ref Young,P.J., Sargent, W.L.W. \& Boksenberg, A. 1982a, ApJ, 252, 10

\ref Young,P.J., Sargent, W.L.W. \& Boksenberg, A. 1982b, ApJS, 48, 455

\bigskip

\newpage

\noindent{\bf Figure captions}

\begin{description}

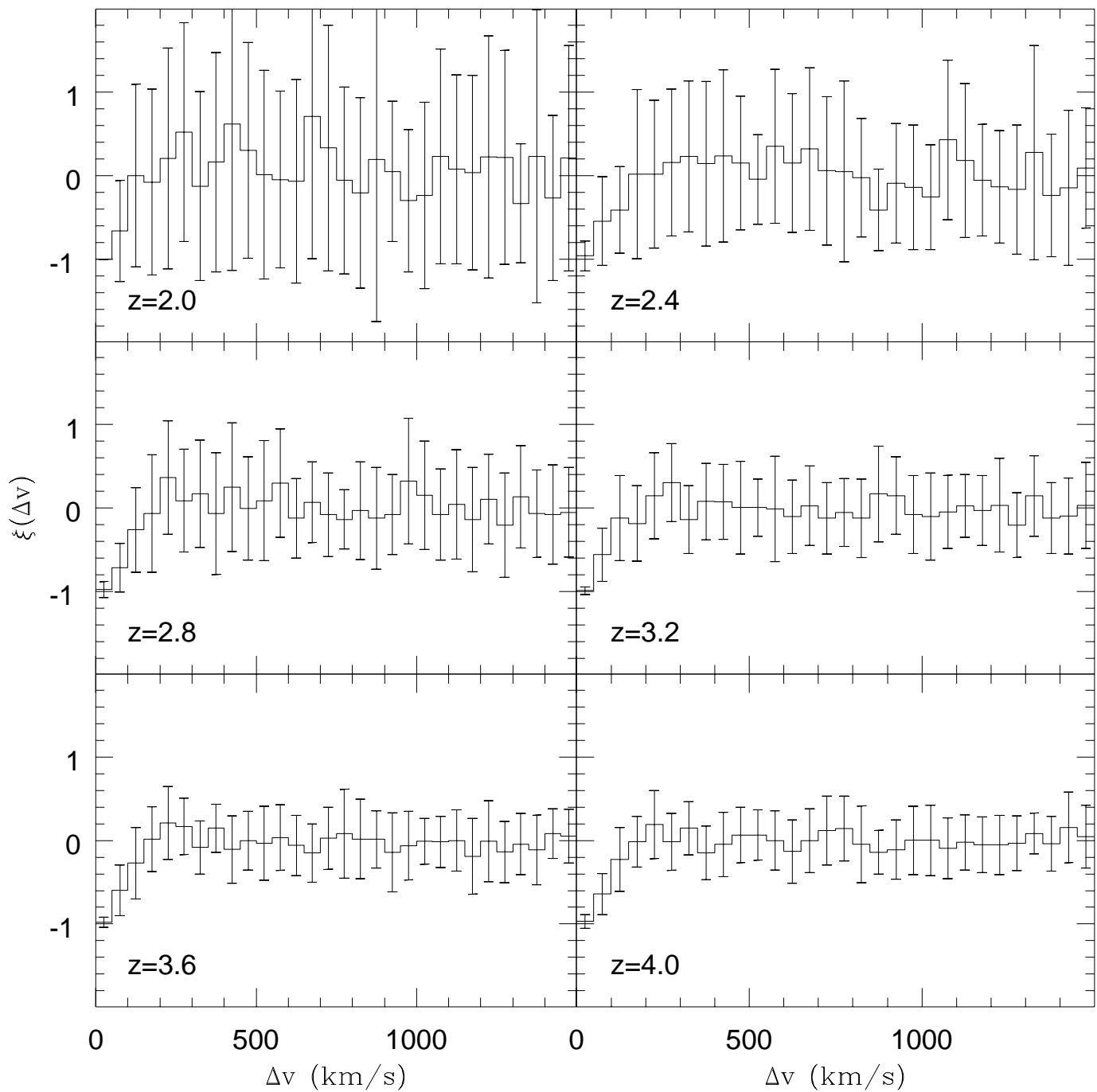
\item[{\bf Figure 1}] A typical two-point correlation function of
Ly$\alpha$ forest lines in velocity space. The sample is taken to be
a simulation of BGF for model of LCDM with $J_{-21}=3.0$. The line
width is taken to be $W_{thr} \geq 0.16$ \AA. The error bars represent
1 $\sigma$.

\item[{\bf Figure 2}] Wavelet reconstruction of density fields. A. the
original (or scale $J$) line distribution of a BGF sample with
$W>0.16 \AA$. B. C. and D. the reconstructed fields for scale $J-1$,
$J-2$ and $J-3$, respectively.

\item[{\bf Figure 3}] The difference of mother function coefficients
between a BGF and random samples at each position $k$. A. B. and C.
Correspond to scales of $J-$, $J-2$ and $J-3$, respectively. The error
bars represent one $\sigma$ around the average of the coefficient
differences given by 100 random sample.

\item[{\bf Figure 4}] The average number $N(>R)$ of clusters identified
from BGF samples, where $R$ is the richness of the clusters in unit
of $\sigma$. A. B. and C. are for scales $J-1$, $J-2$ and $J-3$,
respectively. The error comes from average among 20 BGF samples.

\item[{\bf Figure 5}] The same as Figure 3 but for QSO-0237 forest with
line width $W>0.16 \AA$. A. B. and C. are for scales $J-1$, $J-2$ and
$J-3$.

\item[{\bf Figure 6}] Integral number $N(>R)$ of clusters identified
from the LWT data (1991), where $R$ is the richness of the clusters.
 A. B. and C. are for scales $J-1$, $J-2$ and $J-3$,
respectively.

\item[{\bf Figure 7}] Integral number $N(>R)$ of clusters identified
from a Bechtold data (1994) with $W>0.32\AA$, where $R$ is the richness
of the clusters. A. B. and C. are for scales
$J-1$, $J-2$ and $J-3$, respectively.

\item[{\bf Figure 8}] Integral number $N(>R)$ of clusters identified
from a Bechtold data (1994) with $W>0.16\AA$, where $R$ is the richness
of the clusters. A. B. and C. are for scales
$J-1$, $J-2$ and $J-3$, respectively.

\item[{\bf Figure 9}] The same as Figure 6 but using the Mallat
wavelet as the basis functions of SSD.  Note that numbers $N(>R)$
are very similar to D4 wavelet, although shape of histogram is
slightly different.

\item[{\bf Figure 10}] dN/dz vs (1+z). A. LWT data of $W>0.36\AA$, B.
Bechtold data of $W>0.32 \AA$. The top curves are given by original
Ly$\alpha$ lines. The lower curves are given by clusters on scales
$J-1$, $J-2$ and $J-3$.

\item[{\bf Figure 11}] dN/dz vs (1+z). A. a BGF sample of $W>0.16\AA$,
B. Bechtold data of $W>0.16 \AA$. The top curves are given by original
Ly$\alpha$ lines. The lower curves are given by clusters on scales
$J-1$, $J-2$ and $J-3$. The amplitudes of the BGF curves of $J-1$,
$J-2$ and $J-3$ are much lower than the corresponding amplitudes
of Bechtold data.

\end{description}

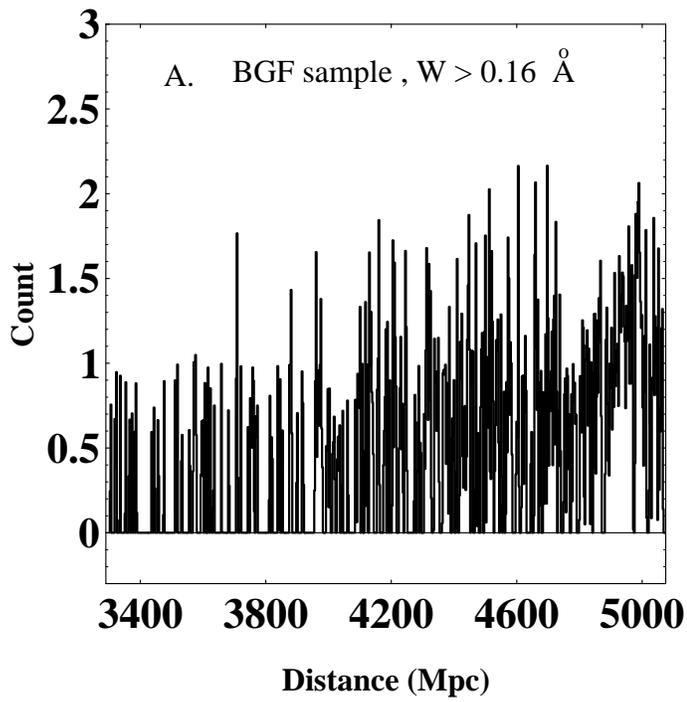
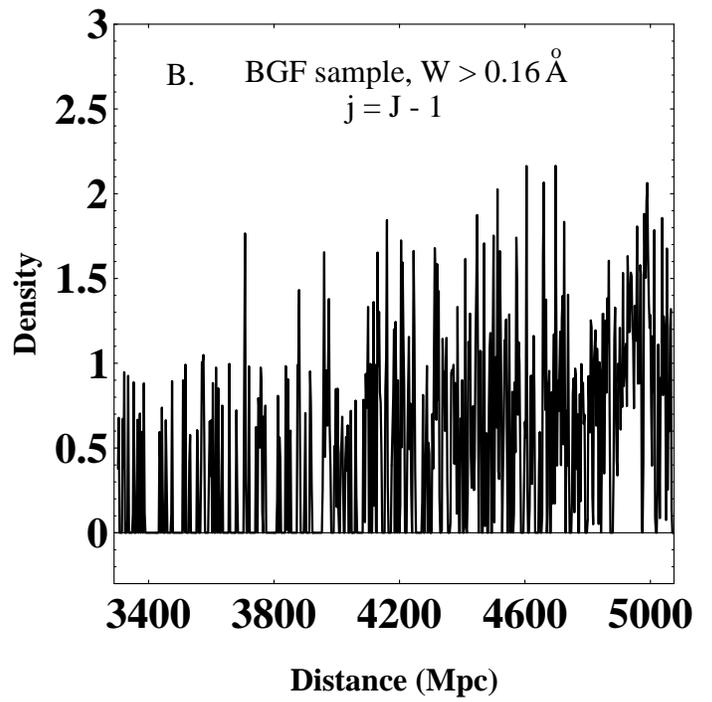
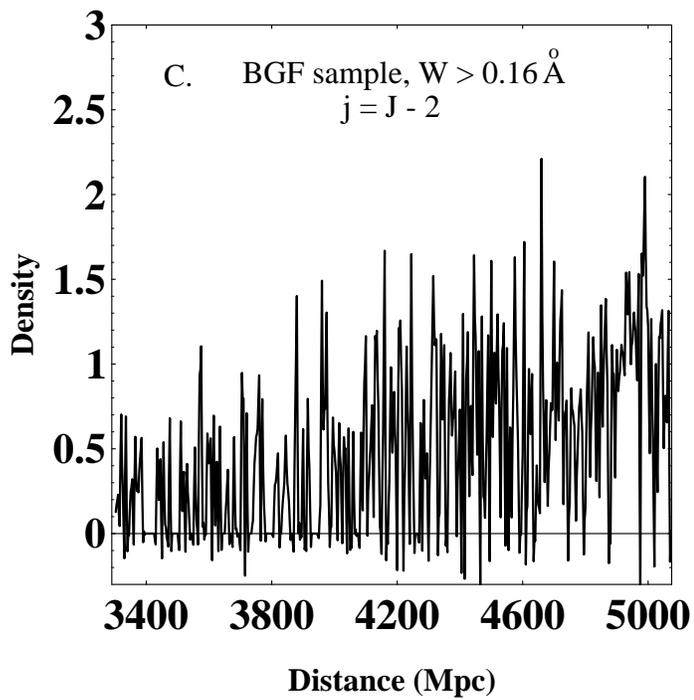
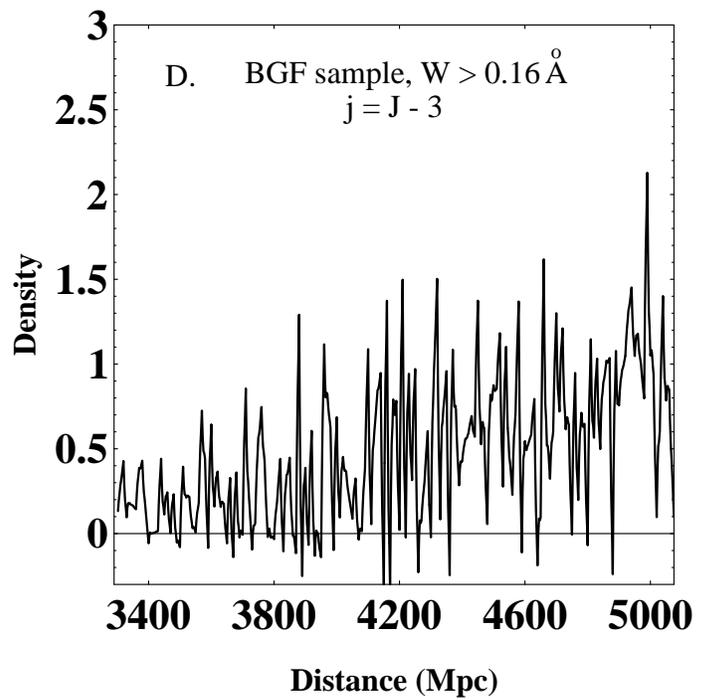
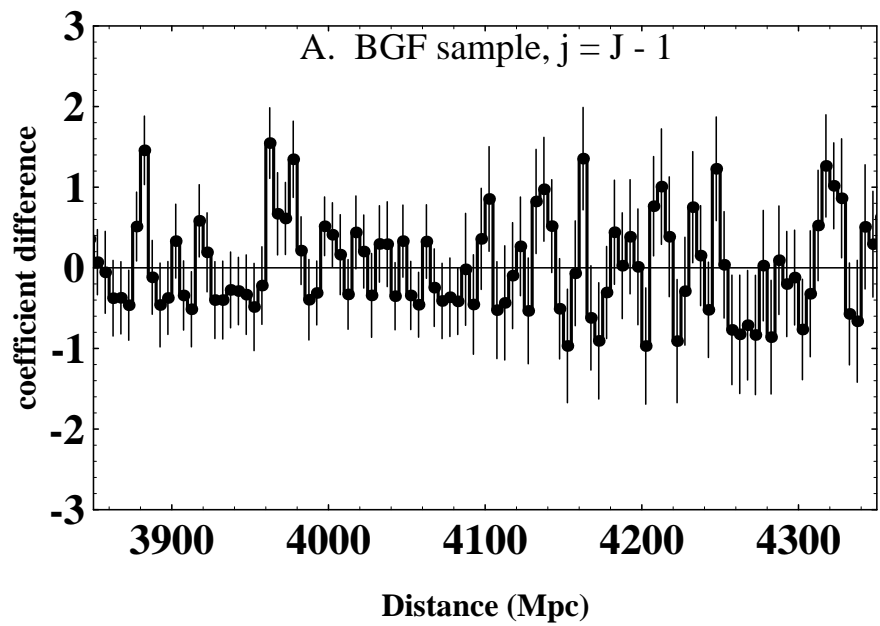
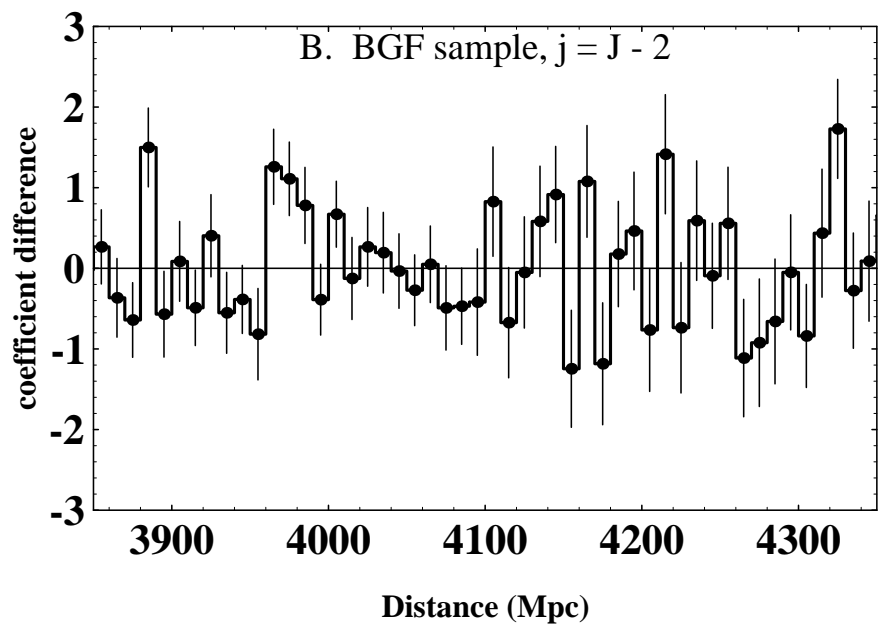
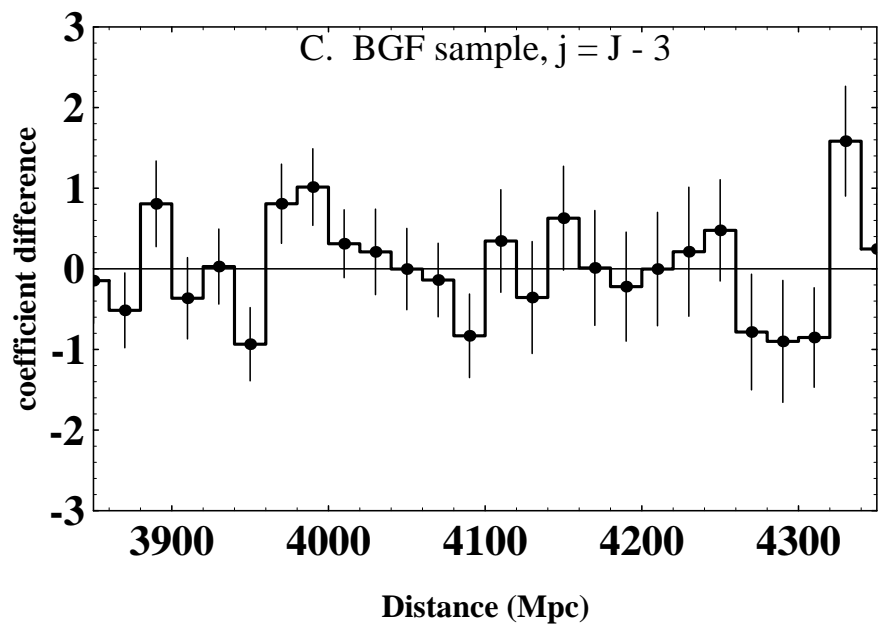
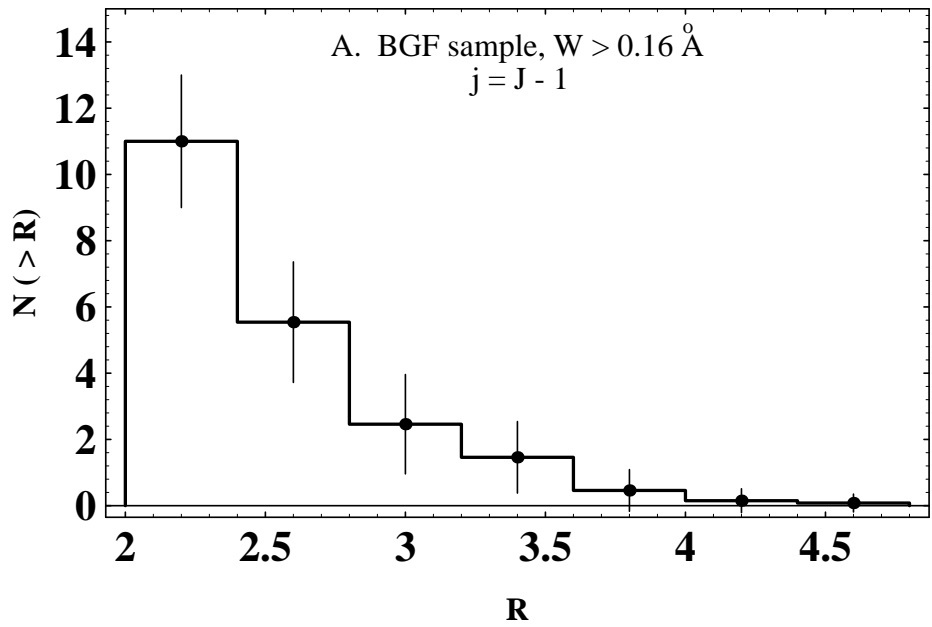
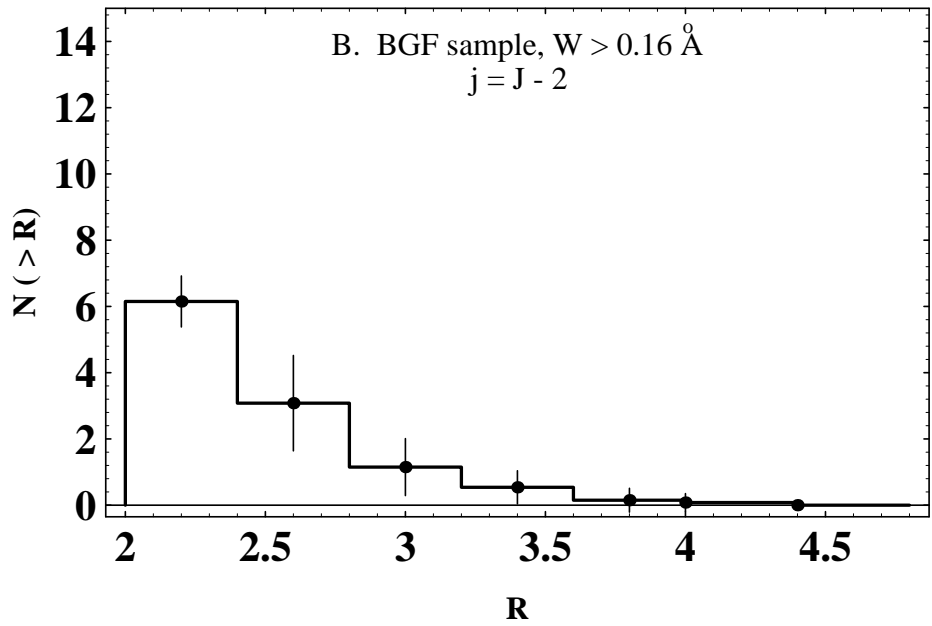
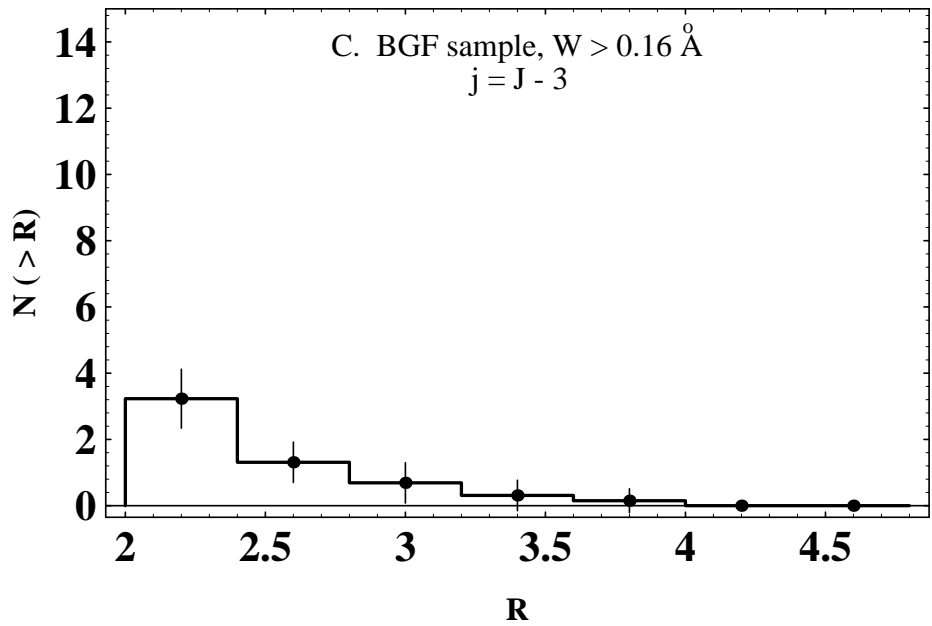
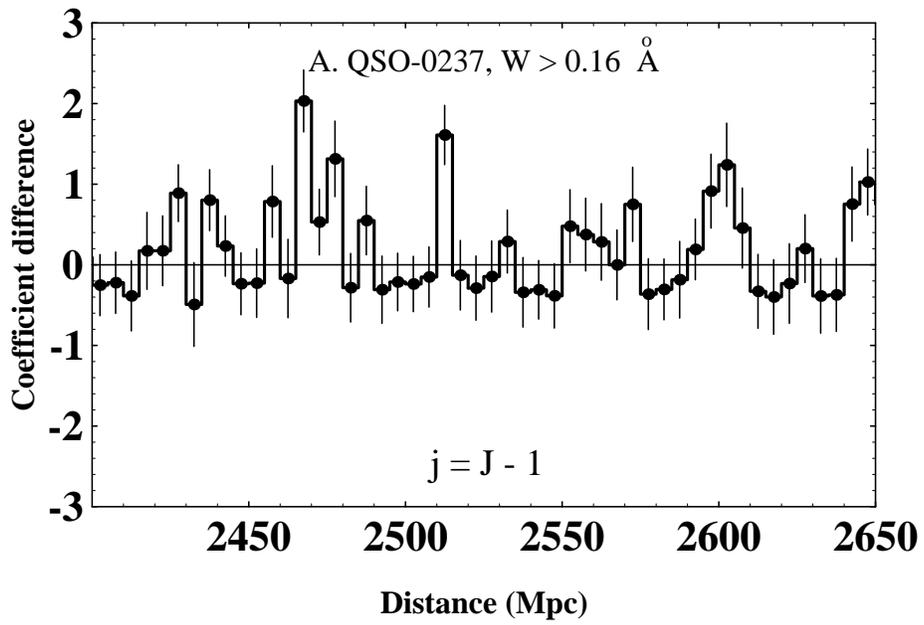
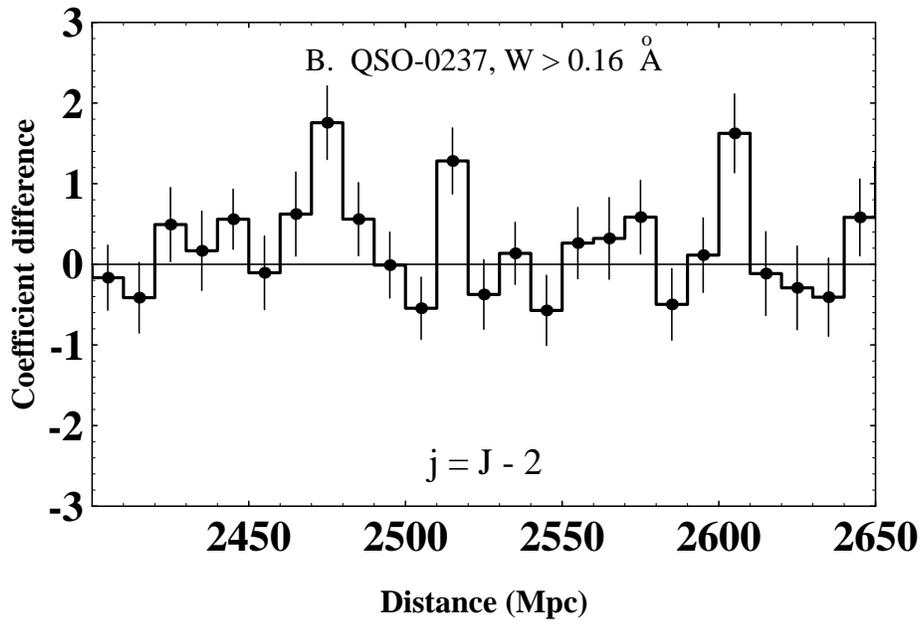
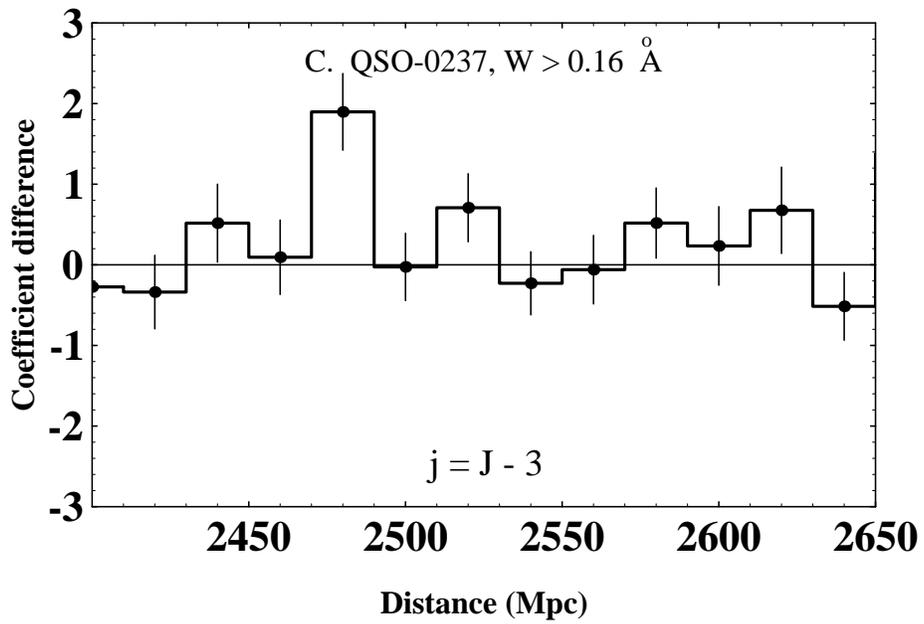
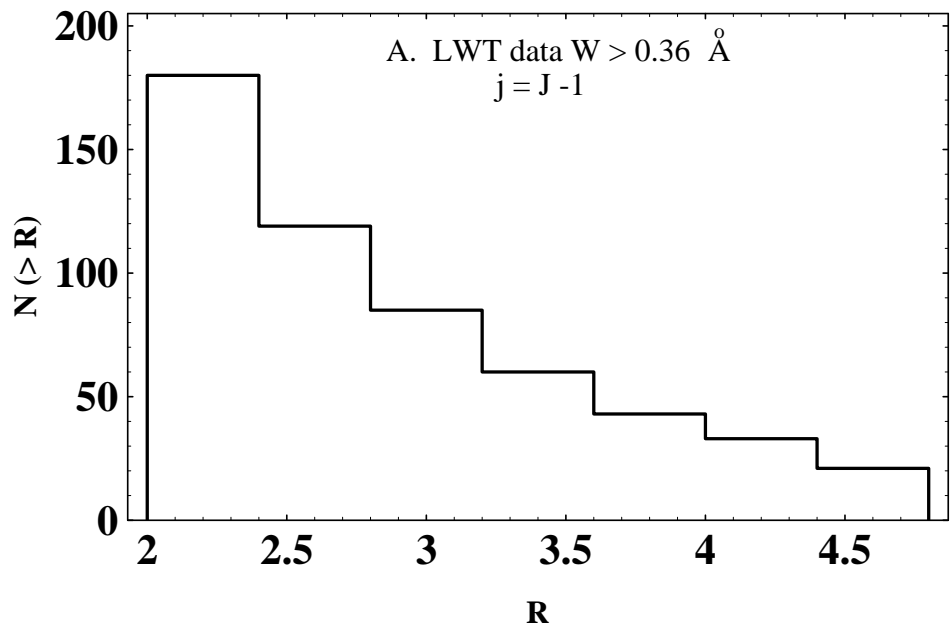
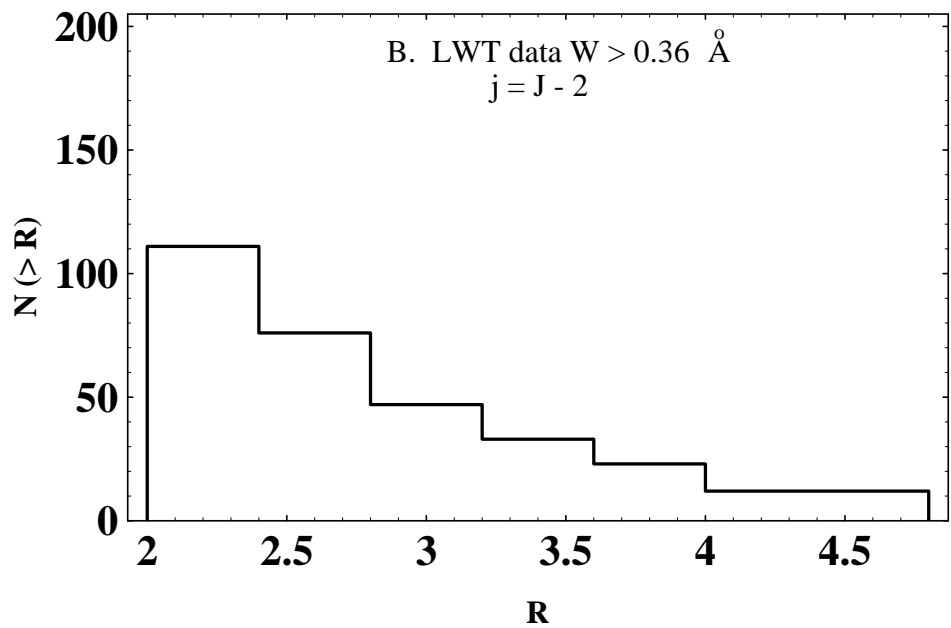
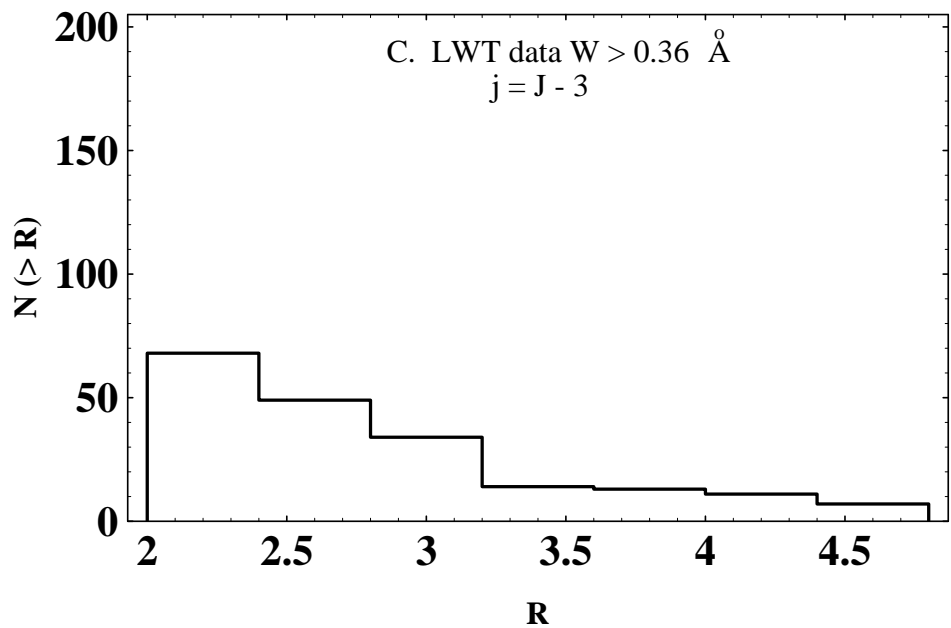
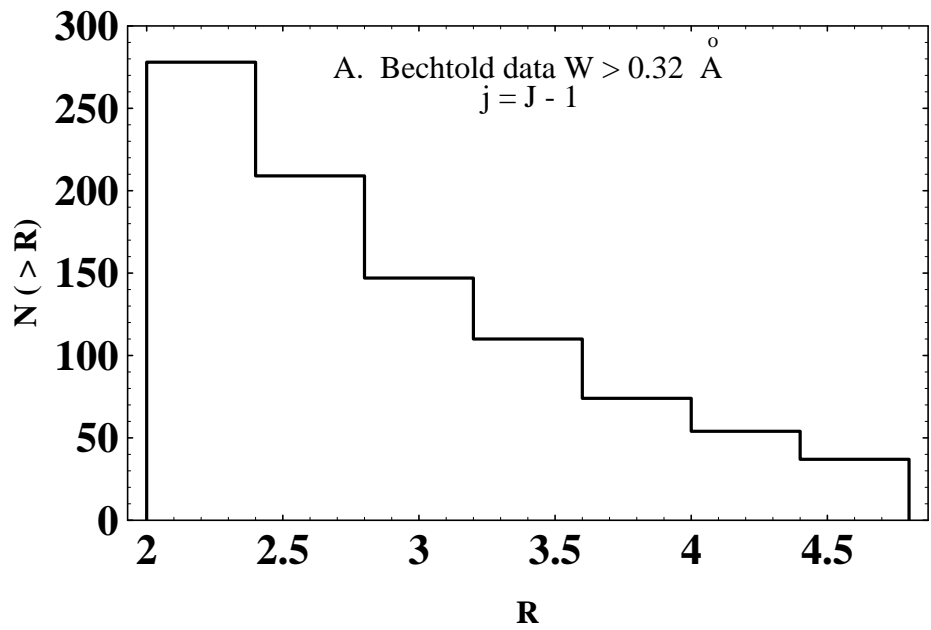
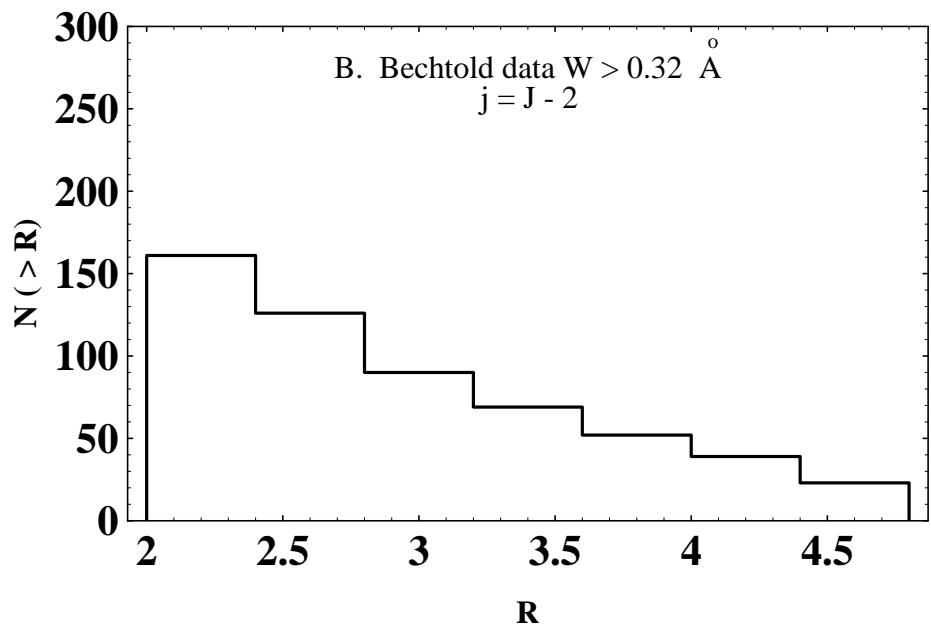
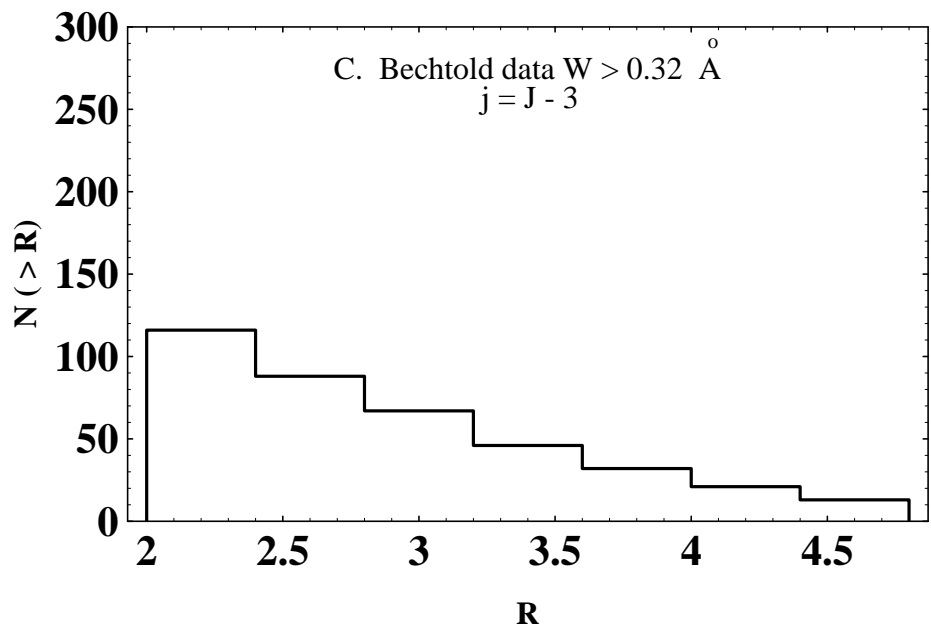
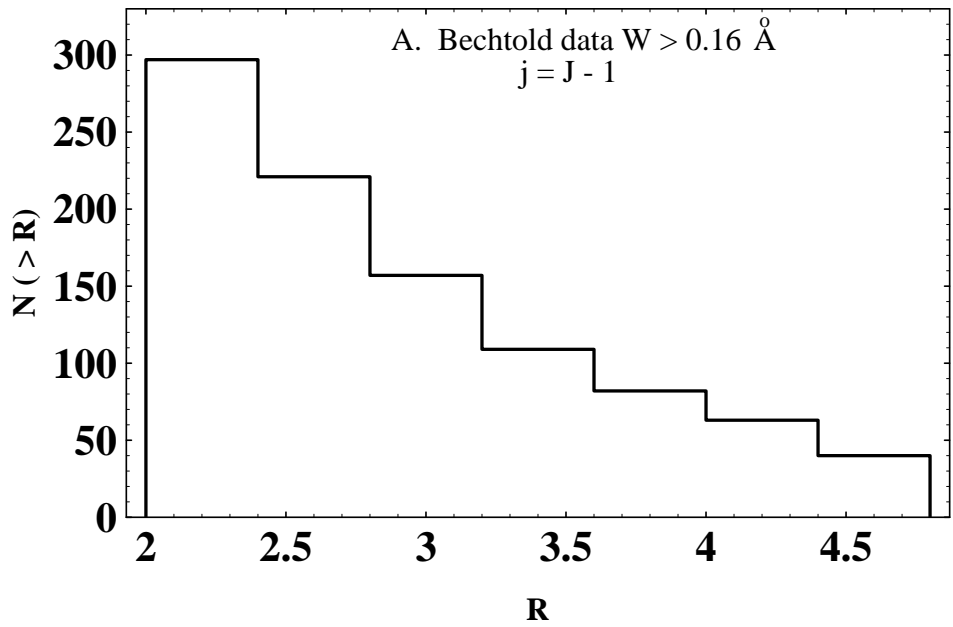
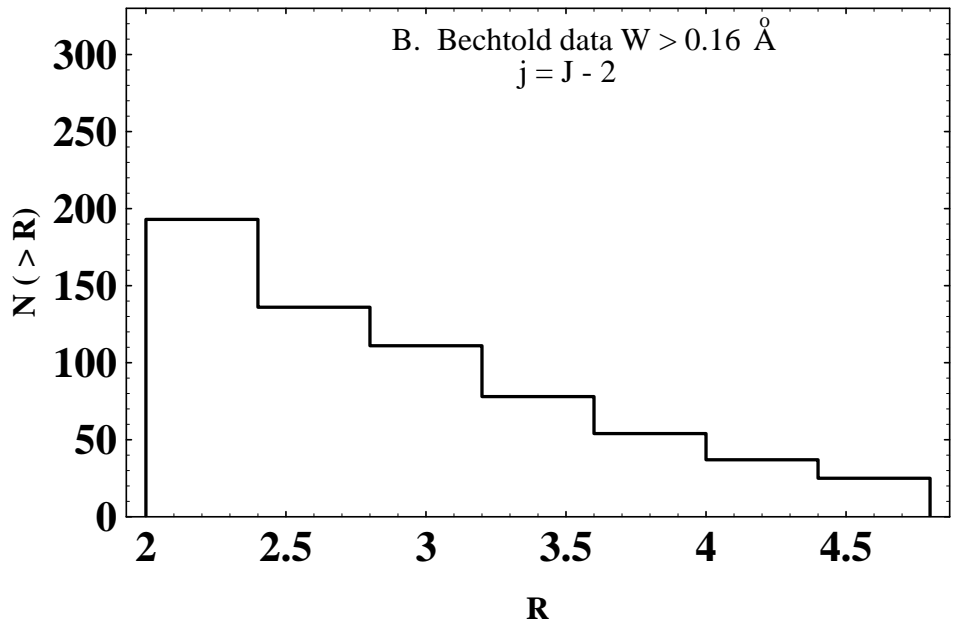
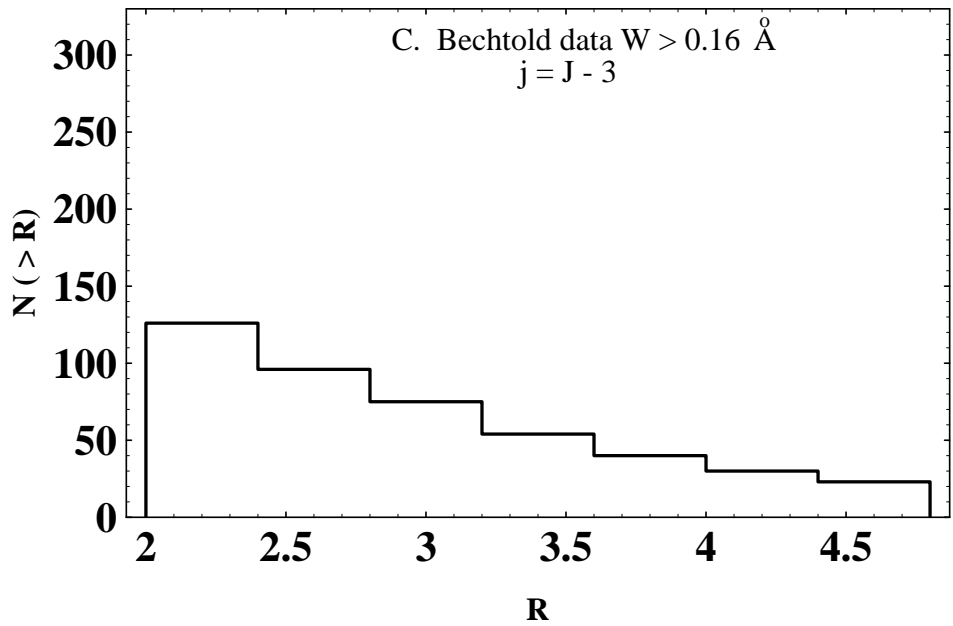
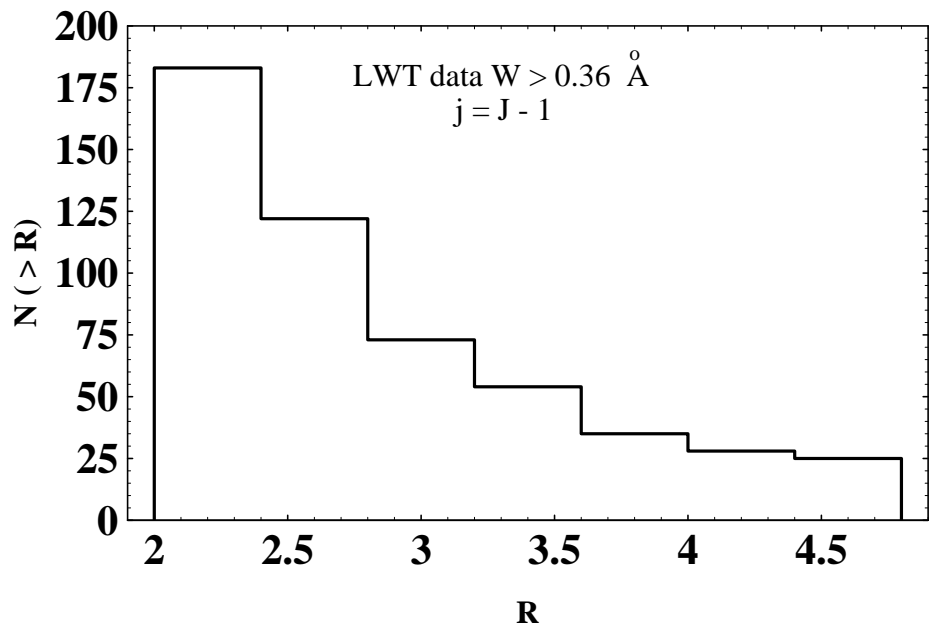
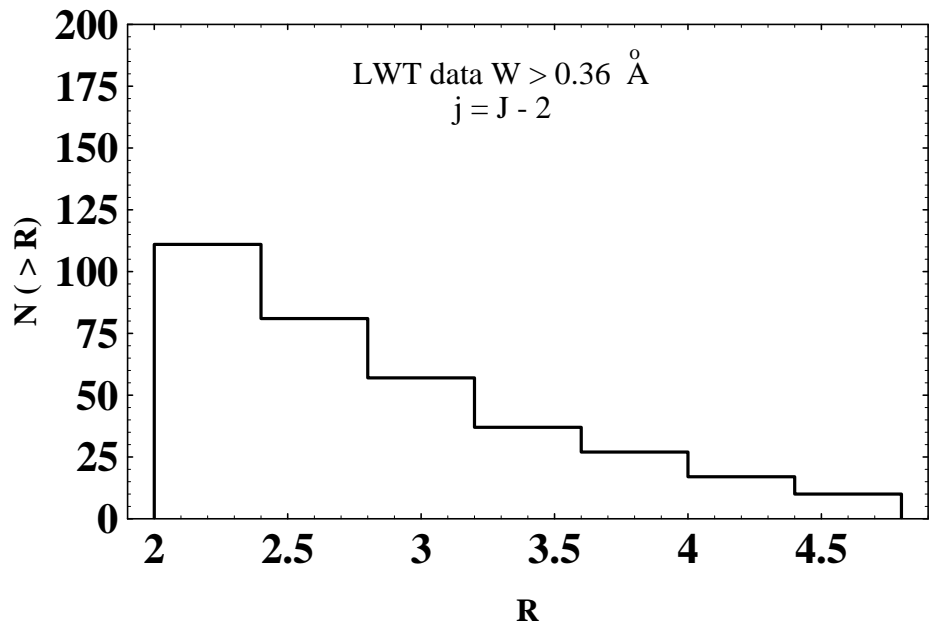
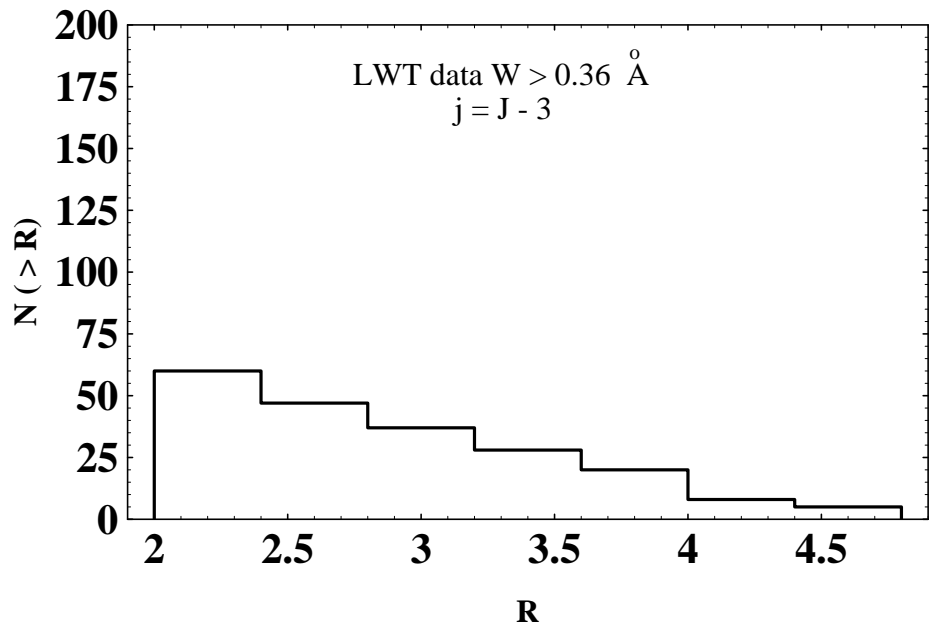
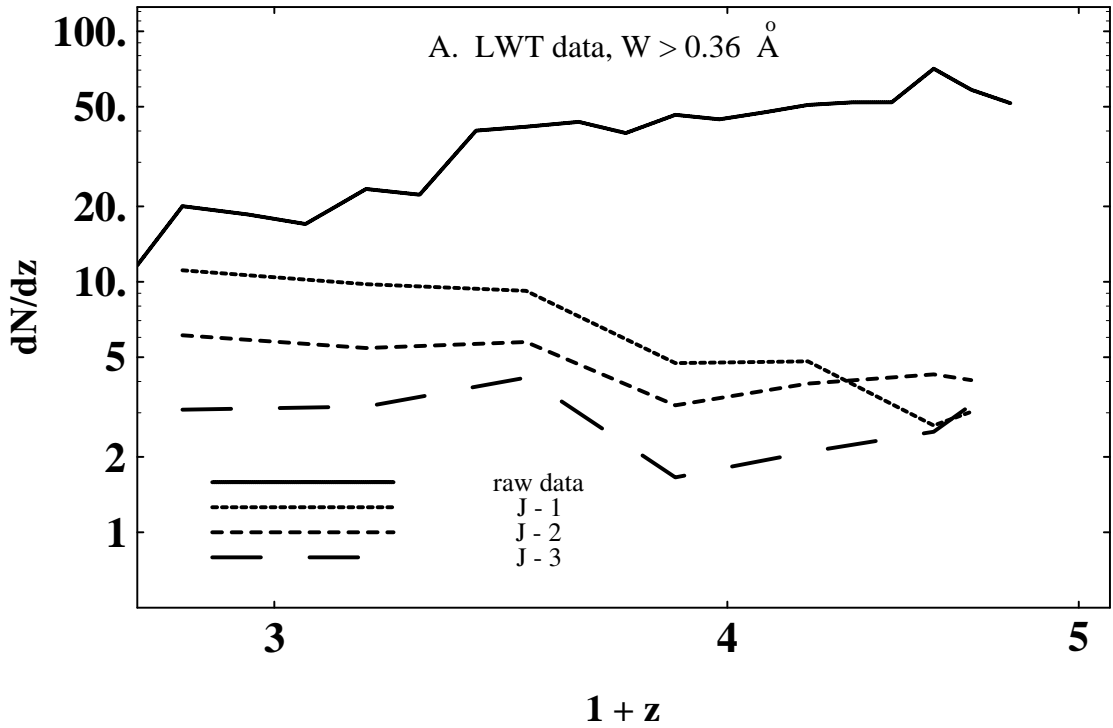
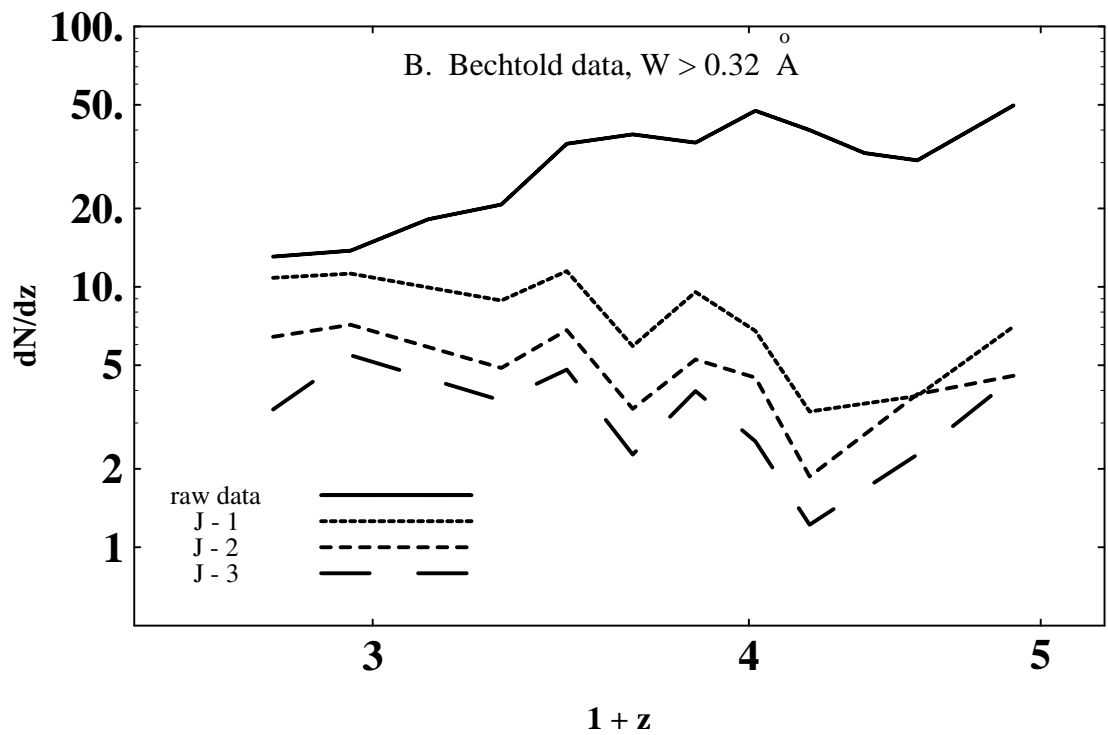
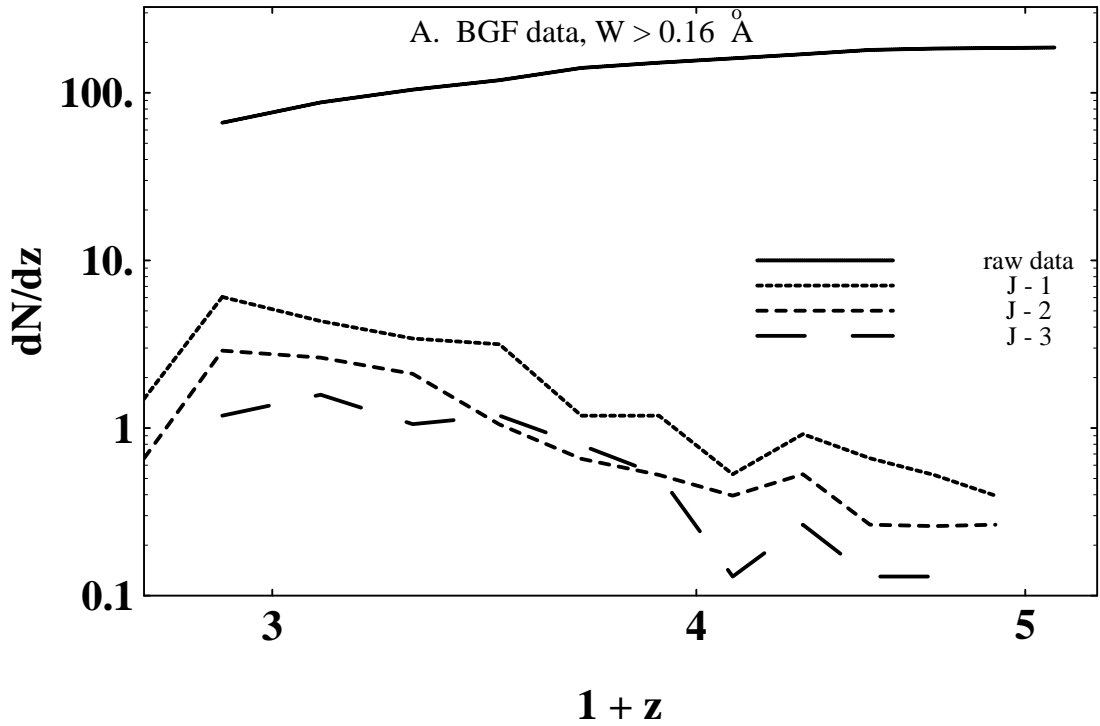
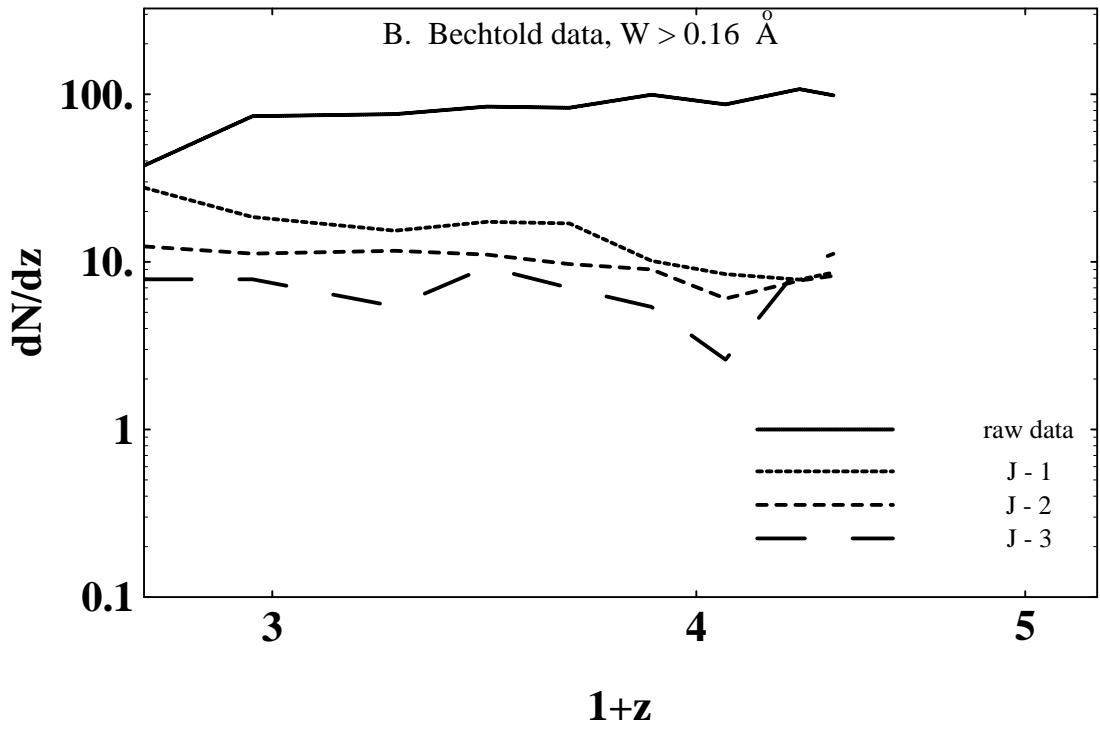
\end{document}